\documentclass{aa}
\usepackage{graphicx}
\usepackage{txfonts}
\usepackage{longtable}
\usepackage{url}
\usepackage{wasysym}
\usepackage{natbib}
\bibpunct{(}{)}{;}{a}{}{,}

\begin{document}
\title{HD~135344B: a young star has reached its rotational limit\thanks{Based on observations collected at the European Southern Observatory, La Silla, Chile (Program ID: 084.A-9016, 085.A-9027, 085.A-9024}}

\author{A.~M\"uller\inst{\ref{inst1},\ref{inst2}}
  \and M.~E.~van~den~Ancker\inst{\ref{inst1}}
  \and R.~Launhardt\inst{\ref{inst2}}
  \and J.~U.~Pott\inst{\ref{inst2}}
  \and D.~Fedele\inst{\ref{inst3}}
  \and Th.~Henning\inst{\ref{inst2}}}
\institute{European Southern Observatory, Karl-Schwarzschild-Str. 2, D-85748 Garching b. M\"unchen, Germany\label{inst1}
  \and Max Planck Institute for Astronomy, K\"onigstuhl 17, D-69117 Heidelberg, Germany\\\email{amueller@mpia.de}\label{inst2}
  \and Johns Hopkins University Dept. of Physics and Astronomy, 3701 San Martin drive Baltimore, MD 21210 USA\label{inst3}
  }
\date{Received date / Accepted date}
\abstract
{}
{
We search for periodic variations in the radial velocity of the young Herbig star HD~135344B with the aim to determine a rotation period.
}
{
We analyzed 44 high-resolution optical spectra taken over a time range of 151 days. The spectra were acquired with FEROS at the 2.2m MPG/ESO telescope in La Silla. The stellar parameters of HD~135344B are determined by fitting synthetic spectra to the stellar spectrum. In order to obtain radial velocity measurements, the stellar spectra have been cross-correlated with a theoretical template computed from determined stellar parameters.
}
{
We report the first direct measurement of the rotation period of a Herbig star from radial-velocity measurements. The rotation period is found to be 0.16~d (3.9~hr), which makes HD~135344B a rapid rotator at or close to its break-up velocity. The rapid rotation could explain some of the properties of the circumstellar environment of HD~135344B such as the presence of an inner disk with properties (composition, inclination), that are significantly different from the outer disk.
}
{}
\keywords{Stars: activity - Stars: atmospheres - Stars: fundamental parameters - Stars: mass-loss - Stars: pre-main sequence - Stars: variables: T Tauri, Herbig Ae/Be}
\maketitle
\section{Introduction}\label{sec:intro}
Herbig Ae/Be (HAeBe) stars \citep{her60,fin84,wat98} are intermediate-mass (2-10 M$_{\odot}$) pre-main sequence stars. They are believed to be higher-mass counterparts of the T~Tauri stars (TTS) and, therefore, fill the parameter space between TTSs and high-mass young stars in addressing the question of star formation as a function of mass \citep{app94}.
\\
Stellar rotation is a crucial parameter for the evolution of angular momentum, magnetic fields, and accretion processes. It is known that HAeBe stars exhibit significantly larger projected rotational velocities, $v\sin i$, than TTSs. Typical $v\sin i$ values for HAeBe stars are in the range of 60 to 225~km~s$^{-1}$ \citep{dav83,fin85,boe95}, whereas most TTSs have $v\sin i$ values of about 10~km~s$^{-1}$ \citep[e.g.][]{wei10}. This result indicates that mechanisms of angular momentum dispersal work much less efficient in HAeBe stars than in TTSs \citep{boe95}. Star-disk locking and subsequent rotational braking might even be absent in HAeBe stars. Only for low-mass HAeBe stars ($M\leq2.6~\mathrm{M}_{\odot}$) \citet{boe95} found indications for loss of angular momentum due to stellar winds.
\\
However, there is no extensive study to determine the rotation periods of HAeBe stars present yet. Only for a few HAeBe stars, rotation periods were found so far. Observed variations of \ion{Ca}{II} K and \ion{Mg}{II} h and k of \object{AB Aur} were interpreted as rotational modulation \citep[e.g.][]{pra86,cat86}. \citet{hub11} detected a rotationally modulated magnetic field of the HAeBe star \object{HD~101412}. In this work we determine the rotation period of the Herbig star \object{HD~135344B} by measuring radial velocity variations using multi-epoch high resolution optical spectra.
\\
The paper is organized as follows: Section~\ref{sec:obs} presents the observations and data reduction. The determination of the stellar parameters and the radial velocity measurements are presented in Sects.~\ref{sec:params} and~\ref{sec:analysis}. A discussion of the results and conclusions can be found in Sects.~\ref{sec:discussion} and~\ref{sec:conclusion}. The Appendix provides a table of measured quantities.
\section{Observations and data reduction}\label{sec:obs}
The observations were carried out in two observing campaigns with the Fiber-fed Extended Range Optical Spectrograph \citep[FEROS;][]{kau99} at the 2.2m MPG/ESO telescope at La Silla Observatory in Chile. FEROS covers the whole optical spectral range from 3600~\AA~to 9200~\AA~and provides a spectral resolution of $\approx$48\,000. The fibre aperture of FEROS is 2~arcsec on sky. A contamination of the spectra due to other stars around HD~135344B can thus be ruled out (Sect.~\ref{sec:params}). In total, 44 spectra of HD~135344B were obtained over a time range of five months between March and July 2010. Depending on the conditions, we observed the star up to four times a night with a separation of two hours between the observations. The average exposure time was 16 minutes per spectrum, which results in an average signal-to-noise ratio (SNR) of 230 at 5500~\AA~of our spectroscopic data set. A set of 40 spectra were obtained using the Object-Calibration mode where one of the two fibers is positioned on the target star and the other fiber is feed with the light of a \element[][]{Th}\element[][]{Ar}+\element[][]{Ne} calibration lamp. This mode allows to monitor and correct for the intrinsic velocity drift of the instrument. Further four spectra were obtained using the Object-Sky mode where the second fiber is pointing to the sky, which allows the subtraction of the sky background from the target spectrum. The reduction of the raw data was performed using the online data reduction pipeline available at the telescope\footnote[1]{\url{http://www.eso.org/sci/facilities/lasilla/instruments/feros/}}. The pipeline does the bias subtraction, flat-fielding, traces and extracts the single echelle orders, applies the wavelength calibration, and corrects for the barycentric motion. For each exposure, it produces 39 individual sub-spectra representing the individual echelle orders as well as one merged spectrum.
\section{Astrophysical parameters of HD~135344B}\label{sec:params}
\begin{figure}
  \resizebox{\hsize}{!}{\includegraphics{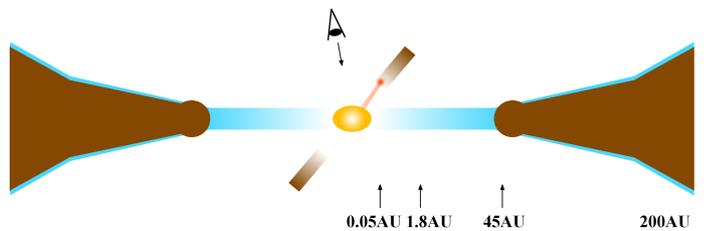}}
  \caption{A pictographic sketch of the HD~135344B system. The star, the dusty disk (brown color) with its large inner gap, the gaseous disk (blue), and an hot spot emerging from the inner rim of the inner disk caused by active accretion (red) are shown.\label{fig:sketch}}
\end{figure}

HD~135344B (SAO~206462) belongs to the Sco~OB2-3 (Upper Centaurus Lupus, UCL) star-forming region, whose center has a distance of $140\pm2$~pc \citep{zee99}. \citet{pre08} also list the individual distances of 81 group members. The median and standard deviation of these values gives $142\pm27$~pc for the distance. Note that this spread reflects the true extent of Sco~OB2-3, rather than observational errors in distances to individual stars. We thus adopt a value of $142\pm27$~pc for the distance of HD~135344B in the following.
\\
\\
HD~135344B is the secondary star of the visual binary system of \object{HD 135344} (SAO~206463). The two components are separated by 21\arcsec~\citep[PA=197$\degr$,][]{mas01}, which translates to a projected separation of 3000~$\pm$~600~AU for the given distance of 142~$\pm$~27~pc. Therefore, a gravitational interaction between the primary and the disk of HD~135344B can be ruled out for this large separation. Using the HST/NICMOS2 camera at $\lambda = 1.6~\mu\mbox{m}$, \citet{aug01} found a close binary system lying 5.8\arcsec~from HD~135344B. A companionship was ruled out by the detection of different proper motions by \citet{gra09} which used the same instrument.
\\
\\
The disk around HD~135344B was subject to numerous studies carried out over a large wavelength range using imaging, spectroscopy and interferometry \citep[e.g.][]{thi01,den05,dou06,bro07,fed08,pon08,gra09,ver10} and revealed a complex structure. An outer radius of 200~AU for the dusty disk was derived by \citet{dou06} at $20.5~\mu\mbox{m}$. From modeling the Spitzer spectrum, \citet{bro07} found a gap in the dusty disk between 0.45~AU and 45~AU. We should note, that \citet{bro07} do not resolve the inner disk but estimated the location of the inner rim of the gap based on potentially ambiguous SED model fits. This is in agreement with the mid-IR observations of \citet{ver10}. One explanation for the presence of a gap in the disk, besides grain growth and photo-evaporation, is the formation of a (sub-)stellar companion which causes a dynamical clearing. \citet{pon08} rule out the presence of a stellar companion around HD~135344B based on their detection of molecular \element{CO} gas ($\lambda = 4.7~\mu\mbox{m}$) at 0.3--15~AU.\\
The value for the disk inclination derived from different observations at different wavelengths varies from 11$\degr$ to 61$\degr$ \citep{den05,dou06,fed08,pon08,gra09}. We should note, however, that different methods probe different spatial scales in the disk. Therefore those different values may not be in contradiction to one another but may reflect the presence of a more complex morphology than assumed in the simple symmetric models by which these values are derived. \citet{gra09} ruled out inclinations greater than 20$\degr$ for the outer disk which is in good agreement with the value of $11\degr\pm2\degr$ measured by \citet{den05}. Therefore, the outer disk is seen almost face-on as it is already indicated by direct imaging \citep{dou06,gra09}.
Interferometric N-band observations of the inner part of the star-disk system revealed the presence of a much higher inclined (53$\degr$-61$\degr$) structure with an estimated inner radius of about 0.05~AU and an outer radius of 1.8~AU \citep{fed08}.
\\
The disk of HD~135344B contains $(2.8\pm1.3)\cdot 10^{-3}~\mbox{M$_{\odot}$}$ of gas and dust \citep{thi01}. The detection of polycyclic aromatic hydrocarbon (PAH) features \citep[e.g.][]{cou95,bro07,fed08} implies a rich chemistry in the vicinity of the star. From J=3-2 \element[][12]{CO} measurements, \citet{den05} derived an inner radius of $\leq$10~AU and 75~$\pm$~5~AU for the outer radius where emitting gas is present.
\\
\\
HD~135344B also shows signs of active mass accretion. This matches the findings of \citet{pot10}, who studied young transitional disk systems with a gap in the circumstellar dust distribution, similar to HD~135344B (Fig.~\ref{fig:sketch}). They systematically found sub-AU scale dust in accreting transitional disk systems. \citet{gar06} derived a value for the mass accretion rate of $\dot{M}_{acc}\approx5.4\cdot10^{-9}~\mbox{M$_{\odot}$~yr$^{-1}$}$ using the \element[][]{Br}$\gamma$ emission line in the NIR. From FUV data, \citet{gra09} concluded that HD~135344B drives no jet, but found indications for a stellar wind leading to a mass-loss rate between $10^{-11}$ to $10^{-8}~\mbox{M$_{\odot}$~yr$^{-1}$}$.
\\
\\
Summarizing all results from the previous observations, the following picture of the disk geometry can be drawn: dust is present within an inner ring from 0.05~AU to 1.8~AU.  This structure is highly inclined compared to the outer disk. There are indications that mass accretion from the inner disk onto the star is still present. Between 1.8~AU and 45~AU, a large gap is present. At 45~AU the outer disk begins and is extended to about 200~AU and is seen almost pole-on. See Figure~\ref{fig:sketch} for a pictographic sketch.
\subsection{Stellar parameters}\label{sec:stellparams}
\begin{figure}
  \resizebox{\hsize}{!}{\includegraphics{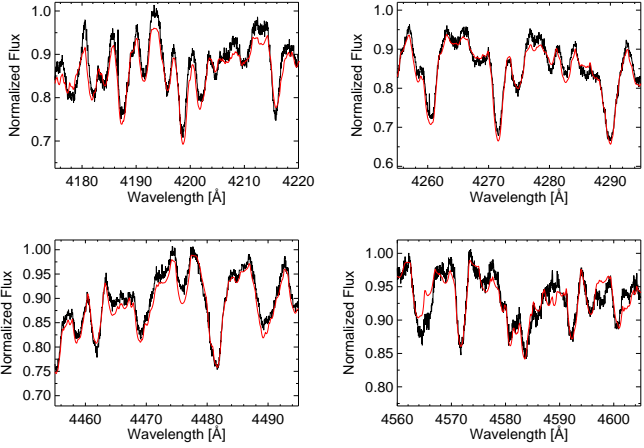}}
  \caption{The spectral windows used for fitting the synthetic spectrum (red line) to the observed stellar spectrum (black line, JD=2455347.63521) of HD~135344B.\label{fig:fitexp}}
\end{figure}

\begin{figure}
  \resizebox{\hsize}{!}{\includegraphics{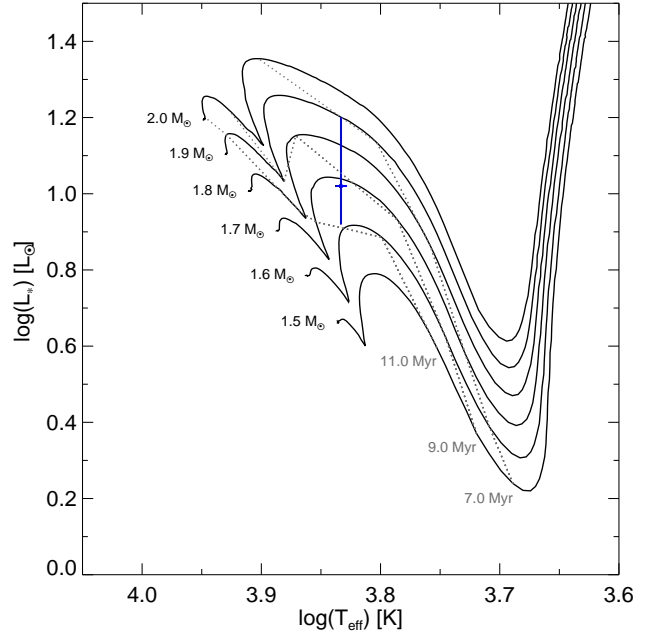}}
  \caption{The position of HD~135344B in an H-R diagram along with the evolutionary tracks of \citet{sie00} for a metal abundance of Z=0.02.\label{fig:hrd}}
\end{figure}

Table~\ref{tbl:stellparam} lists the derived stellar parameters of HD~135344B. The luminosity, $L_{\star}$, was computed from integrating the total flux under the SED. $T_\mathrm{eff}$, $\log g$, and $v\sin i_{\star}$ were computed by a self-developed tool for retrieving stellar parameters of Herbig stars based on fitting synthetic spectra to the observed stellar spectrum. The computation of the synthetic spectrum was carried out using {\sc spectrum}\footnote[2]{\url{http://www.phys.appstate.edu/spectrum/spectrum.html}} \citep{gra94} with the {\sc atlas9} atmosphere models \citep{cas04}. For a detailed description of this procedure we refer to \citet[][in preparation]{mue11}. Figure~\ref{fig:fitexp} shows a part of an stellar spectrum of HD~135344B (black line, observed at JD=2455347.63521) which is well fitted by a single star synthetic spectrum (red line).\\
From the position of HD~135344B in the H-R diagram (Fig.~\ref{fig:hrd}), we obtained the stellar mass and radius and its age. The derived parameters are well in agreement with literature values, e.g., \citet{pla08,man06,boe05,dun97}. The measured effective temperature is comparable to that of an F3Ve to F4Ve main-sequence star and agrees with the value found by \citet{dun97}. 
\begin{table}
  \caption{Stellar parameters of HD~135344B. \label{tbl:stellparam}}
  \centering
  \begin{tabular}{lr}
  \hline\hline
  Parameter & Value \\
  \hline
  $distance$ & 142 $\pm$ 27~pc\\
  $\log L_{\star}$ & 1.02$^{+0.18}_{-0.10}$~L$_{\odot}$\\
  $T_\mathrm{eff}$ & 6810 $\pm$ 80~K\\
  $\log g$ & 4.4 $\pm$ 0.1 cm~s$^{-2}$ \\
  $M_{\star}$ & 1.7$^{+0.2}_{-0.1}$ M$_{\odot}$ \\
  $age$ & 9 $\pm$ 2 Myr\\
  $R_{\star}$ & 1.4 $\pm$ 0.25 R$_{\odot}$ \\
  $<RV_{\star}>$ & 2.5 $\pm$ 1.5 km~s$^{-1}$ \\
  $v\sin i_{\star}$ & 82.5 $\pm$ 2.9 km~s$^{-1}$ \\
  \hline
  \end{tabular}
  \tablefoot{
    $<RV_{\star}>$ represents the mean value of the observed RVs. The error does not present the accuracy of the determined $RV_{\star}$ but reflects the scatter on $RV_{\star}$ around the mean value.}
\end{table}
\section{Data analysis and results}\label{sec:analysis}
\subsection{Radial velocity measurements}\label{sec:rv}
The measurement of the stellar $RV$ has been done by cross-correlating the stellar spectrum with a template spectrum. The template was a synthetic spectrum representing the fitted stellar parameters of HD~135344B (Table~\ref{tbl:stellparam}). We measure a projected rotational velocity, $v\sin i_{\star}$, of 82.5~km~s$^{-1}$ (Sect.~\ref{sec:stellparams}). Therefore, all stellar spectral lines are highly broadened and blended, which makes a determination of $RV_{\star}$ difficult since the accuracy is limited to some 100~m~s$^{-1}$. For the cross-correlation, only the spectral range between 4000~\AA~to 7875~\AA~was considered because the instrument efficiency drops significantly outside this spectral range. In addition, areas with strong emission lines, telluric lines, and Balmer lines were carefully marked out. The resulting cross-correlation function was fitted by a Gaussian function. The position of the center of the Gaussian yields $RV_{\star}$. After the individual $RV_{\star}$ values were derived separately for each echelle order, the median value and standard deviation of $RV_{\star}$ were computed. We applied to the $RV_{\star}$ values a 1-$\sigma$ clipping to remove anomalous velocity values, similar to e.g. \citet{bar05,jef07}. The deficient $RV_{\star}$ estimates, rejected by this $\sigma$ clipping, do not follow a simple Gaussian noise statistics but are far off the median $RV_{\star}$, which indicates individual, systematic biases of the $RV_{\star}$ estimate in the rejected orders. The final $RV_{\star}$ were then computed based on averaging ten to twelve remaining individual orders. The error of the final $RV_{\star}$ is the standard deviation of the mean. All measured $RV_{\star}$ values are listed in Table~\ref{tbl:rv} provided in the appendix. Figure~\ref{fig:RVall} shows the measured RVs for all 44 spectra with a mean $RV_{\star}$ value of 2.5~km~s$^{-1}$, indicated by the gray horizontal dashed line. Peak-to-peak variations of up to 2.9~km~s$^{-1}$ are present.\\
\begin{figure}
  \resizebox{\hsize}{!}{\includegraphics{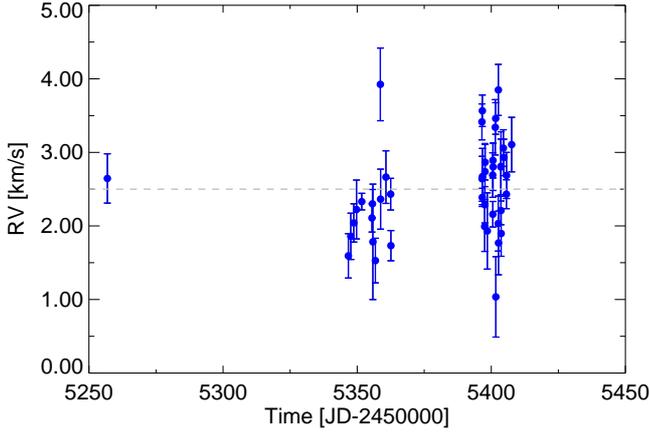}}
  \caption{Measured RVs for HD~135344B. The horizontal gray dashed line represents the mean value of the measurements.\label{fig:RVall}}
\end{figure}
\subsection{Radial velocity variations}\label{sec:RVvariations}
\begin{figure}
  \resizebox{\hsize}{!}{\includegraphics{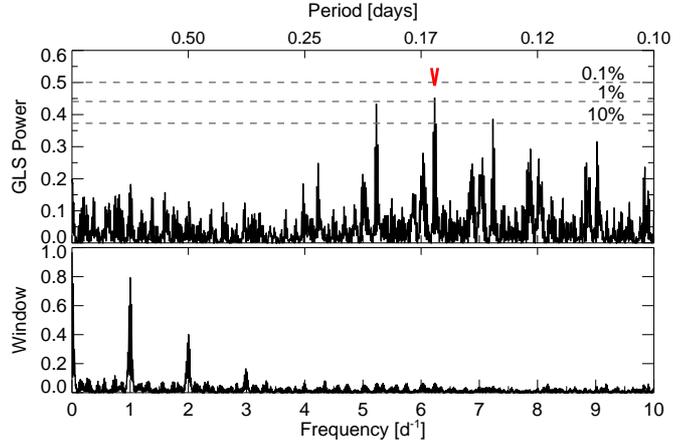}}
  \caption{GLS periodogram (upper plot) and window function (lower plot) of the $RV_{\star}$ data. The significant period at 0.16~d is marked by the red arrow. FAP thresholds for 0.1\%, 1\%, and 10\% are indicated by the horizontal dashed lines. The data are plotted over frequency but the upper x-axis gives the corresponding period. Note that periods greater than 1~d are in the frequency range between 0 and 1~d$^{-1}$.\label{fig:GLSRV}}
\end{figure}

\begin{figure}
  \resizebox{\hsize}{!}{\includegraphics{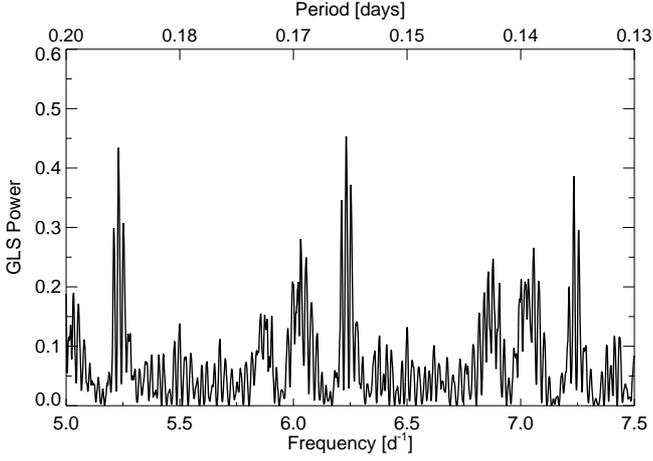}}
  \caption{Close-up view of the GLS centered around the most significant peak at 0.16~d. Two less significant peaks at 5.230~d$^{-1}$ and 7.234~d$^{-1}$ are present in addition and are aliases caused by the sidereal day. All three peaks show side lobes with a difference of 0.021~d$^{-1}$ in frequency, caused by the uneven sampling of our data.\label{fig:GLSRVzoom}}
\end{figure}

\begin{figure}
  \resizebox{\hsize}{!}{\includegraphics{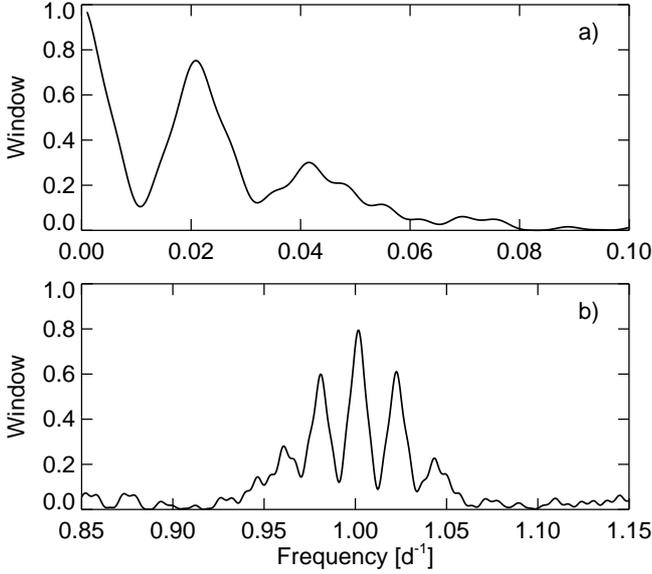}}
  \caption{Close-up views of the window function for the significant peaks at $<$0.006~d$^{-1}$, 0.021~d$^{-1}$, and 1.002~d$^{-1}$.\label{fig:GLSwinzoom}}
\end{figure}

\begin{figure}
  \resizebox{\hsize}{!}{\includegraphics{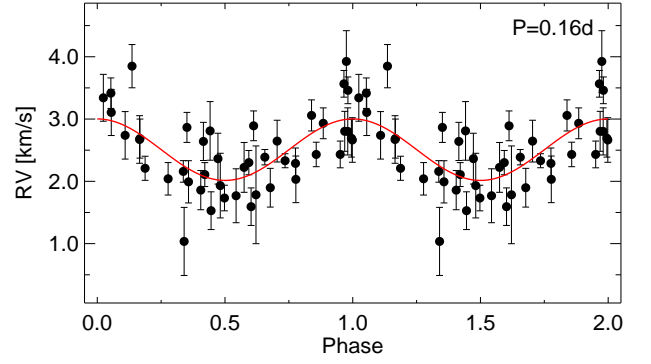}}
  \caption{Phase-folded plot of the $RV_{\star}$ data. The solid line is a sinusoidal fit with a period of $0.16045\pm2\cdot10^{-5}$~d and an amplitude of $493\pm86$~m~s$^{-1}$.\label{fig:RVphase}}
\end{figure}

\begin{figure}
  \resizebox{\hsize}{!}{\includegraphics{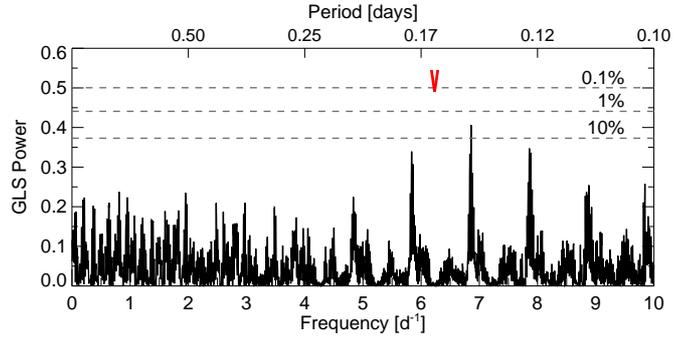}}
  \caption{GLS periodogram of the residual $RV_{\star}$ data after subtraction of the sinusoidal fit. The red arrow marks the position of the 0.16~d period which is no longer present, as well as its corresponding aliases. The new peak with the highest power at 0.14~d has a FAP of 3.4\% and is therefore not significant. \label{fig:GLSRVres}}
\end{figure}

In order to identify periodicities present in the $RV_{\star}$ data, we computed the generalized Lomb-Scargle (GLS) periodogram \citep{zec09} and its window function (Fig.~\ref{fig:GLSRV}). The GLS exhibits a strong peak at a period of 0.16045~d (in the following we use 0.16~d), marked by the red arrow. Its false-alarm probability (FAP) is 0.64\%. In addition, there are several other peaks in the GLS present which are less or not significant compared to the 0.16~d period. This makes the interpretation of a periodogram ambiguous. In order to identify alias frequencies caused by uneven sampled data in time, the window function has to be considered. Significant peaks in the window function can produce aliases in the GLS which are separated from the true peak by the frequency difference. Figure~\ref{fig:GLSRVzoom} shows a close-up view of the GLS, centered around the 0.16~d period (or at the frequency of 6.232~d$^{-1}$ respectively). The two strongest other, but less significant peaks are located at 5.230~d$^{-1}$ and 7.234~d$^{-1}$. The difference of both frequencies to 6.232~d$^{-1}$ is 1.002~d$^{-1}$, which corresponds to the sidereal day. Figure~\ref{fig:GLSwinzoom}~b) shows a close-up view of the window function centered around the significant frequency 1.0~d$^{-1}$. The highest peak is at a frequency of 1.002~d$^{-1}$. Therefore, we can conclude that the peaks to the left and to the right of the 0.16~d period in the GLS are aliases. Each of the three peaks in Fig.~\ref{fig:GLSRVzoom} show side lobes which are separated by 0.021~d$^{-1}$ for all cases. From the close-up view of the window function (Fig.~\ref{fig:GLSwinzoom}~a), we can identify a significant frequency present at 0.021~d$^{-1}$ which is exactly the measured frequency difference between the side lobes of three main peaks. The 0.021~d$^{-1}$ frequency corresponds to a period of about 48~d. By looking at the distribution of observations in Fig.~\ref{fig:RVall}, we can identify three separated data sets (the first data set consists of the first single data point only). The time difference between the first and the second data set is about 100~d, and the difference between the second and the third data set is about 50~d. This explains the observed aliases at this frequency difference in the GLS and the presence of a significant peak in the window function. The observations cover a time range of about 150~d. Therefore, the window function (Fig.~\ref{fig:GLSwinzoom}~a) shows a significant peak for frequencies smaller than 0.006~d$^{-1}$ (or periods greater than 150~d respectively), but they are not visible as side lobes of the main peaks in the GLS. In addition, we subtracted a sinusoidal fit from the $RV_{\star}$ data and computed a GLS periodogram of the residuals (Fig.~\ref{fig:GLSRVres}) in order to verify that the 0.16~d period is present in the $RV_{\star}$ data. The GLS periodogram of the residuals clearly shows that the peak at 0.16~d as well as its corresponding aliases disappeared. There are no further significant peaks at other periods present. From this analysis, we consider the 0.16~d period as significant and as the only real period present in the observed $RV_{\star}$ data.
\\
Figure~\ref{fig:RVphase} shows the phase-folded $RV_{\star}$ data for the 0.16~d period. The solid line represents a sinusoidal fit to the data with a period of $0.16045\pm2\cdot10^{-5}$~d and an amplitude of $493\pm86$~m~s$^{-1}$. The results of the fit are presented in Table~\ref{tbl:resultssinfitRV}.
\begin{table}
\caption{Results of the sinusoidal fit to the $RV_{\star}$ data. \label{tbl:resultssinfitRV}}
\centering
  \begin{tabular}{lc}
  \hline\hline
  Parameter & Value \\
  \hline
  $\chi^{2}_{\mathrm{red}}$ & 2.5 \\
  \emph{rms} & 478~m~s$^{-1}$ \\
  \emph{period} & $0.16045\pm2\cdot10^{-5}$~d \\
  \emph{amplitude} & $493\pm86$~m~s$^{-1}$ \\
  \emph{phase} (to JD$_{\mathrm{min}}$) & $0.95\pm0.03$ \\
  \emph{offset} & $2508\pm61$~m~s$^{-1}$ \\
  \hline
  \end{tabular}
\end{table}
\subsection{Bisector analysis}\label{sec:bisector}
\begin{figure}
  \resizebox{\hsize}{!}{\includegraphics{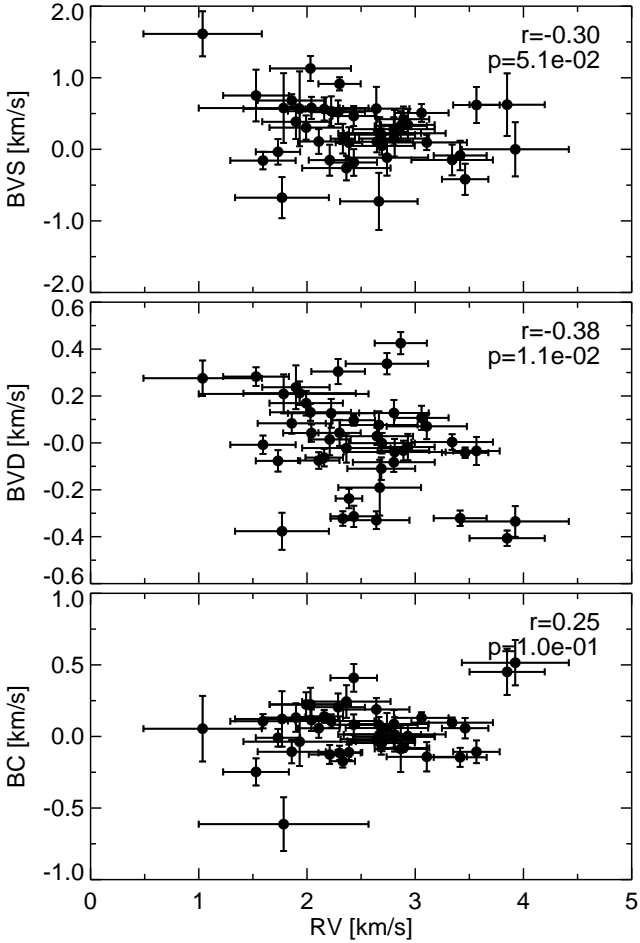}}
  \caption{Bisector quantities vs. $RV_{\star}$. For each plot the linear correlation coefficient $r$ and the probability $p$ that the data are linearly uncorrelated are computed \citep{bev03}. There is no linear correlation between the bisector quantities and $RV_{\star}$ present.\label{fig:BS}}
\end{figure}

Stellar activity like cod and hot spots, granulation and pulsation can mimic significant RV variations that can be periodic due to stellar rotation. In case of a cold (i.e. dark) spot, the line profile becomes asymmetric because of the reduced stellar light. Stellar rotation leads to a modulation of the asymmetry and causes RV variations depending on the stellar inclination and latitude of the spot. The analysis of the line profile (in the following bisector) is routinely applied in the search for extrasolar planets when the RV technique is used in order to identify if the observed periodic RV variations are caused by stellar activity or due to a companion \citep[e.g.][]{que01}. A correlation between $RV_{\star}$ and bisector velocity span would indicate the the $RV_{\star}$ variation is caused by line shape asymmetries, i.e. by rotational modulation, e.g., due to a spot. The shape of the computed cross-correlation function (CCF) represents the mean line profile of the selected lines in the observed spectra. We measured the bisector velocity span (BVS), bisector displacement (BVD), and bisector curvature (BC) for all our spectra following the definition by \citet{pov01}. To construct the bisector, the middle points of the horizontal segments, which connect a point on the left and on the right side of the CCF at the same flux level, are computed. The bisector gets divided into three zones. The BVS is simply the difference between the upper and lower zone. The BC is defined as the difference between the upper and lower bisector spans. To derive the BVD the average of the three bisector velocity zones is computed. For the computation of the individual quantities we chose heights of 30\%, 50\%, and 85\% of the bisector. Similar to the $RV_{\star}$ measurements, we measured BVS, BVD, and BC for each computed CCF. The final bisector values were derived by averaging the single values. The errors of BVS, BVD, and BC are the standard deviations of the means. In Figure~\ref{fig:BS}, we plotted all three bisector quantities against RV. The computed linear correlation coefficients $r$ are noted in the upper right corner of each plot. For none of the three quantities a significant linear correlation is present. Only a weak trend might be present for BVS and BD. However, the sensitivity of the bisector method decreases rapidly when it comes to low stellar inclinations \citep[e.g.][]{des07}. For HD~135344B it can be assumed that we see the star almost pole-on (Sect.~\ref{sec:discussion}) and therefore we would not expect a strong correlation between BVS and RV.
\subsection{\element[][]{H}$\alpha$ measurements}\label{sec:halpha}
\begin{figure}
  \resizebox{\hsize}{!}{\includegraphics{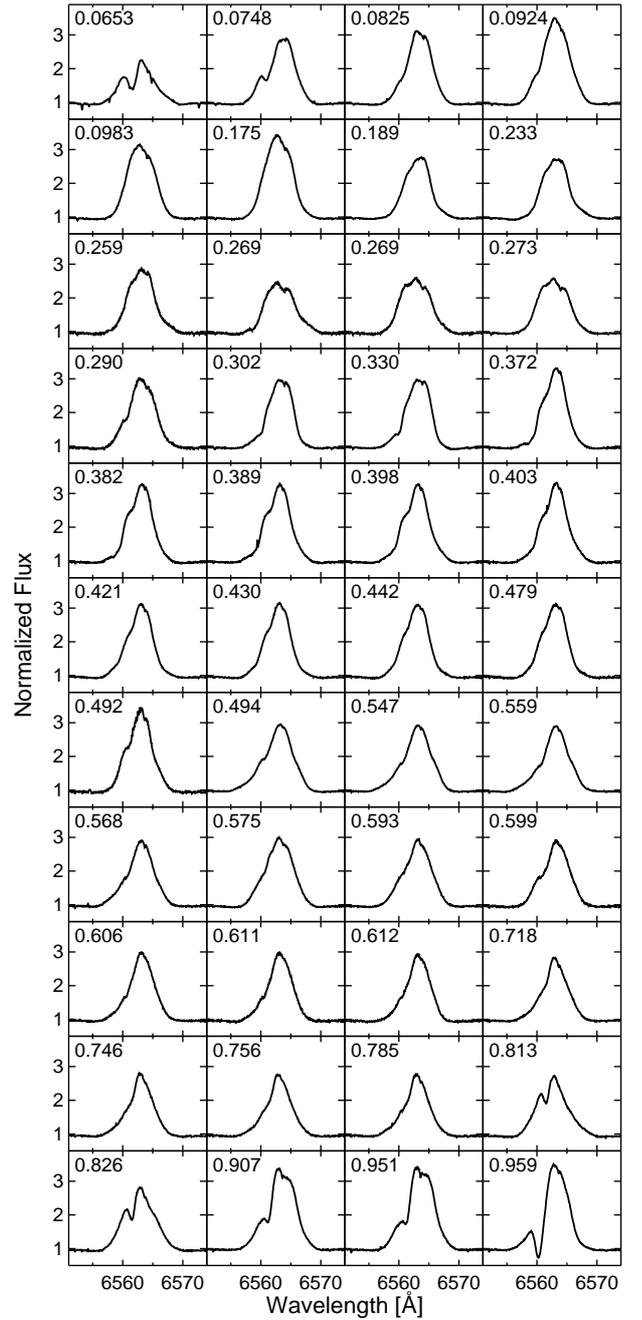}}
  \caption{\element[][]{H}$\alpha$ line profiles of HD~135344B. The plots are ordered with respect to their phase value according to a 5.77~d period, which is shown in the upper left corner of each plot.\label{fig:Ha}}
\end{figure}

\begin{figure}
  \resizebox{\hsize}{!}{\includegraphics{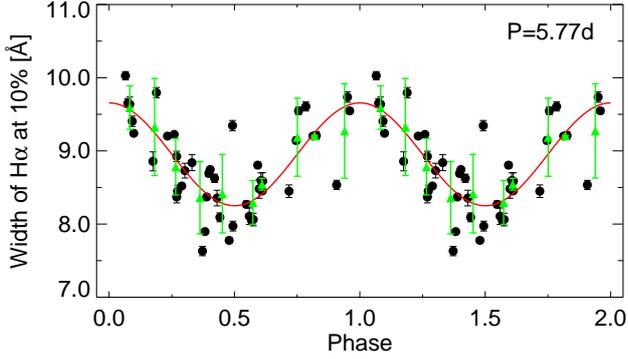}}
  \caption{Phase-folded plot of the $W_{10}(\element[][]{H}\alpha)$ data. The solid line is a sinusoidal fit with a period of $5.77\pm0.02$~d and an amplitude of $0.70\pm0.08$~\AA. The green triangles are the average values computed by binning the phase in 0.1 wide bins. Their error bars are derived by computing the standard deviation of the corresponding $W_{10}(\element[][]{H}\alpha)$ values in each bin.\label{fig:10wHa_phase}}
\end{figure}

The \element[][]{H}$\alpha$ line profile of HD~135344B undergoes rapid and significant changes. \citet{fin84} classified the \element[][]{H}$\alpha$ line profile of HAeBe stars into three categories: single-peak, double-peak, and P~Cygni profiles. The \element[][]{H}$\alpha$ line profile of HD~135344B adopt each of these types at different times. Figure~\ref{fig:Ha} shows the \element[][]{H}$\alpha$ line profiles of our spectroscopic data set of HD~135344B ordered with respect to their phase value according to a 5.77~d period. A double-peaked profile indicates accretion. A blue-shifted absorption (P~Cygni profile) indicates an outflow of material. %
We measured the equivalent width, $EW(\element[][]{H}\alpha)$, of HD~135344B in our FEROS spectra. The measured values are listed in Table~\ref{tbl:rv} provided in the appendix. For the measurement of $EW(\element[][]{H}\alpha)$, we selected several distinct wavelength intervals on both sides of the \element[][]{H}$\alpha$ line and used a 2nd order polynom to describe the continuum. This was repeated four times for different intervals. The deviation of all four measurements defines our error. For $EW(\element[][]{H}\alpha)$, we obtained a mean value of -10~\AA. The GLS periodogram of the $EW(\element[][]{H}\alpha)$ values did not show a significant period.
\\
We also measured the full width of \element[][]{H}$\alpha$ at 10\% height, $W_{10}(\element[][]{H}\alpha)$, which is frequently used for TTSs in order to distinguish between chromospheric activity and accretion of circumstellar material \citep{whi03,nat04,jay06}. We applied the same method for normalizing the spectra as for $EW(\element[][]{H}\alpha)$. We measured a mean value of 9.8~\AA~for $W_{10}(\element[][]{H}\alpha)$, which corresponds to an accretion rate of $3\cdot10^{-9}~\mbox{M$_{\odot}$~yr$^{-1}$}$ if Eq.~5 in \citet{nat04} is applied. The measured values of $W_{10}(\element[][]{H}\alpha)$ are listed in Table~\ref{tbl:rv} provided in the appendix.
The periodogram of $W_{10}(\element[][]{H}\alpha)$ shows a significant period at 5.77~d with a FAP of $1.9\cdot10^{-6}$. A phase-folded plot for the $W_{10}(\element[][]{H}\alpha)$ data (black data points) and the sinusoidal fit (red line) is shown in Fig.~\ref{fig:10wHa_phase}. Assuming Keplerian rotation of the circumstellar gas, this period corresponds to a radial distance of $\approx0.07$~AU. The error bars of $W_{10}(\element[][]{H}\alpha)$ are the statistical errors of the single measurements and reflect the precision of the individual measurements. These error bars are most likely overestimating the accuracy of each measurement, since the distance of the data points to the sinusoidal fit often exceeds the error bars significantly. A possible reason for an accuracy bias of individual $W_{10}(\element[][]{H}\alpha)$ measurements can be that the \element[][]{H}$\alpha$ line can form in different locations. Also, the formation can be caused by different processes, e.g. accretion column, accretion shock on the stellar surface, disk surface, and wind. Therefore, we see a significant scatter of the measured $W_{10}(\element[][]{H}\alpha)$ values around the sinusoidal fit, the level of which exceeds the precision of the individual measurements. However, the bias of each individual measurement statistically spread around the fit. We binned the phase in 0.1 wide bins and averaged the $W_{10}(\element[][]{H}\alpha)$ values lying in the corresponding bin. The shown error is the standard deviation of the values in each bin. These values are overplotted in Fig.~\ref{fig:10wHa_phase} with green triangles. The green error bars are now reflecting both the precision and accuracy of the $W_{10}(\element[][]{H}\alpha)$ measurements and are well described by the sinusoidal fit.
\section{Discussion}\label{sec:discussion}
The observed 0.16~d period might be caused by rotational modulation of one or more spot(s) on the stellar surface. Adopting the stellar radius of $1.4\pm0.25$~R$_{\odot}$ and $v\sin i_{\star}=82.5\pm2.9$~km~s$^{-1}$ from Table~\ref{tbl:stellparam}, the inclination of the stellar rotation axis is $i_{\star}=11\degr~\pm~2\degr$. We thus see the star almost pole-on. This could explain the weak correlation between the BVS and $RV_{\star}$ (Sect.~\ref{sec:bisector}). The disk of HD~135344B is also found to be almost face-on with inclination measurements of $\lesssim 20\degr$ \citep{gra09} and $11\degr\pm2\degr$ by \citet{den05}, i.e., the orientation of the stellar rotation axis, derived from the assumption that the 0.16~d $RV_{\star}$ period represents the stellar rotation period, would be almost perpendicular to the mid-plane of the outer disk. This supports the hypothesis that the 0.16~d period is caused by photospheric effects.\\
Another approach to verify the origin of the 0.16~d period is to apply a Keplerian fit to the data. Under the assumption of a circular orbit (eccentricity is set to 0) and a period of 0.16~d, the semi-major axis is 7$\cdot$10$^{-3}$~AU, which corresponds to 1.5~R$_{\odot}$. For HD~135344B, we found the stellar radius to be 1.4~$\pm$~0.25~R$_{\odot}$ (Table~\ref{tbl:stellparam}), i.e, the semi-major axis would be in the range of the stellar radius. This shows that the companion hypothesis is not plausible, and thus favors a photospheric interpretation. Thus, the rotational period of 0.16~d or 3.9~hours makes HD~135344B an extremely rapid rotator.
\subsection{HD~135344B is rotating close to break-up velocity}
The break-up velocity $v_c = \sqrt{GM_{\star}/R_{\star}}$ of the star, where $G$ is the gravitational constant, has a value of $480~\pm~60$~km~s$^{-1}$. Using the value of $i_{\star}=11\degr~\pm~2\degr$ derived from the stellar parameters leads to a true rotational velocity of $432\pm81$~km~s$^{-1}$. Therefore, HD~135344B is rotating at or close to the break-up velocity at its equator.
\\
\citet{auf06} measured the polar and equatorial radius of \object{Vega} interferometrically which rotates at 91\% of its break-up velocity. They measured a difference of 20\% between equatorial and polar radius and found the temperature gradient to be 2250~K. For HD~135344B, similar effects are expected. Because the star is seen almost pole-on, we might overestimate its luminosity and thus its mass and temperature. The angular stellar diameter of HD~135344B is about 100~$\mu$as and not resolvable by current stellar interferometric facilities which prevents a direct confirmation and the determination of its true stellar parameters.
\\
The continuous accretion of circumstellar material onto the star could have caused the spin-up of the stellar rotation up to $v_c$. HD~135344B has only a very weak magnetic field \citep{hub09}. Therefore, the star is probably decoupled from its disk, which prevents the star from magnetic braking and loosing spin. In contrast, for the majority of TTSs only moderate $v\sin i_{\star}$ values are observed which indicates a disk-braking mechanism during the accretion phase \citep[e.g.][]{wei10}.
\subsection{Possible origins of the observed periodicity}
\subsubsection{The von Zeipel effect}
Because of its rapid rotation, the star should show an oblate shape of its stellar atmosphere due to the centrifugal force. This results in a redistribution of the flux which is proportional to the local surface gravity \citep{zei24}, i.e., the equatorial zones are darker with a lower $T_{eff}$ than the polar zones. This phenomenon is known as gravity-darkening and is symmetric with respect to the equator. In order to produce a periodic $RV_{\star}$ signal, i.e. mimic a stellar spot, the flux and temperature gradient has to be off-centered from the rotational axis in order to produce a Doppler shift, i.e. the von Zeipel effect cannot account for the observed periodicity of the spectra. \citet{smi74} and \citet{cla00} showed that for differentially rotating radiative stellar atmospheres, the flux over the stellar surface varies stronger than the surface gravity, which could produce asymmetric flux distributions.
\subsubsection{Stellar spots}
We computed different spot configurations (temperature difference of the spot with respect to the photosphere $\Delta T_{\mathrm{spot}}$, filling factor $f_{\mathrm{spot}}$, latitude of the spot $\Theta$) taking the determined stellar parameters into account.\\ 
The effects of star spots on photometry and RV$_{\star}$ were numerically derived with a relatively simple model. A sphere with radius $R$, unit surface brightness $I$, and one or more round spots with spot filling factor $f_{\rm spot}$\ at specified latitude and longitude was projected on a 2D Cartesian grid with 200\,$\times$\,200 pixels. The relative spot brightness at the considered wavelength was calculated from the ratio of effective photospheric temperature to spot temperature, assuming black-body radiation. To account for opacity effects on the stellar surface, a linear limb darkening law with coefficients adopted from \citet{cla03} was applied. The inclination of the rotation axis can be arbitrarily chosen. Solid-body rotation was assumed; differential rotation was not considered. Each cell of the star image was thus assigned a brightness and a radial velocity value. The surface brightness distribution at each phase of the stellar rotation (in steps of 1~deg) was then integrated to obtain the total brightness and the position of the photo center (1$^{\rm st}$\ moment). The 1$^{\rm st}$\ moment of the radial velocity distribution, weighted with the surface brightness distribution, was adopted as a proxy for the the mean radial velocity. Absorption line shapes were not taken into account.
\\
The two extreme scenarios which can explain the $RV_{\star}$ variations are: 1.) dark spot, $\Delta T_{\mathrm{spot}}=-1350~\mathrm{K}$, $\Theta=40\degr$, $f_{\mathrm{spot}}=5\%$, 2.) bright spot, $\Delta T_{\mathrm{spot}}=2000~\mathrm{K}$, $\Theta=85\degr$,  $f_{\mathrm{spot}}=56\%$. The large covering fraction of the latter model may more closely resemble the von Zeipel effect than traditional star spots. So far, there are no photometric monitoring data for HD~135344B available which would set additional constraints on the possible spot configurations.
\\
We conclude that both explanations for the origins of the observed $RV_{\star}$ variations, such as stellar spots or the flux and temperature gradient between the pole and equator, can adequately explain the data.
\\
\\
Due to the accretion process, the $W_{10}(\element[][]{H}\alpha)$ variations may trace a hot spot from an accretion funnel flow which has its origin at the inner edge of the inner disk. A periodic variation of 5.77~d is found for $W_{10}(\element[][]{H}\alpha)$ (Sect.~\ref{sec:halpha}). This period corresponds to a Keplerian radius of $\approx$0.07~AU that is close to the inner edge of the disk estimated to 0.05~AU by \citet{fed08}, given the uncertainty. A similar effect was observed for $EW(\element[][]{H}\alpha)$ variations for the \object{TW Hya} system by \citet{set08}.
\\
Several studies have shown that the disk lifetimes of young stars are expected to be $\sim$5 to $\sim$10~Myr \citep[e.g.][]{hai01,bou06,jay06,agu06,fed10}. At this age, stars lose their inner dusty disk and accretion stops. This seems to coincide with the dissipation of gas in the inner disk, too \citep{jay06}. In contrast to that, HD~135344B shows still accretion signatures, an inner dusty disk with an inclination, which is much greater than that of the outer disk, as well as a gas-rich inner environment at an age of $9\pm2$~Myr. A possible explanation might be that HD~135344B is losing parts of its outer atmosphere due to the rapid rotation at or close to its break-up velocity and feeding the inner part of the disk with its own material, which then accretes back onto the star.
\\
A possible stellar or sub-stellar companion located in the gap of the disk suggested by literature \citep[e.g.][]{gra09} cannot be confirmed or ruled out with our measurements because our spectroscopic data set covers only five months in total and the uncertainties in the $RV_{\star}$ measurements are of the order of 300~m~s$^{-1}$.
\section{Conclusions}\label{sec:conclusion}
In this paper we presented the first direct measurement of the stellar rotational period of a Herbig star. The period was derived from $RV_{\star}$ measurements by cross-correlating optical high-resolution spectra with a synthetic spectrum. The rotational period of HD~135344B was found to be 0.16~d. In addition, we determined new reliable stellar parameters for HD~135344B, independent from previous literature values, and demonstrated the possibility to measure the $RV_{\star}$ of a Herbig star with a high $v\sin i_{\star}$ value. From the measured stellar parameters, we were able to estimate the stellar inclination to a value of $11\degr\pm2\degr$. With these data, we concluded that HD~135344B is rotating at or close to its break-up velocity and feeding its stellar vicinity with gas and dust. This is the first observational evidence that a young intermediate-mass star can rotate close to its break-up velocity.
\\
The direct determination of the stellar rotational period by $RV_{\star}$ measurements may be applicable also to other Herbig Ae/Be stars. A larger sample could therefore reveal the evolution of angular momentum, the presence of star-disk coupling mechanisms and the presence possible magnetic fields of these young intermediate-mass stars.
\\
We also found variations in the $W_{10}(\element[][]{H}\alpha)$ measurements with a period of 5.77~d, corresponding to a Keplerian radius close to the inner edge of the inner disk of HD~135344B, which could be a hot spot caused by an accretion funnel flow.
\begin{acknowledgements}
 We thank our colleagues T. Anguita, R. Lachaume, and S. Protopapa for carrying out the FEROS observations. We thank B. Acke for providing us an additional spectrum of HD~135344B. This research has made use of NASA's Astrophysics Data System Bibliographic Services.
\end{acknowledgements}

\bibliographystyle{aa}
\bibliography{refs}
\appendix
\section{$RV_{\star}$ and \element[][]{H}$\alpha$ measurements}
\begin{table*}
  \caption{Measured $RV_{\star}$, $EW(\element[][]{H}\alpha)$, $W_{10}(\element[][]{H}\alpha)$, and their corresponding uncertainties.\label{tbl:rv}}
  \centering
  \begin{tabular}{lcccccc}
  \hline\hline
  Time & $RV_{\star}$ & $\sigma_{RV_{\star}}$ & $EW(\element[][]{H}\alpha)$ & $\sigma_{EW(\element[][]{H}\alpha)}$ & $W_{10}(\element[][]{H}\alpha)$ & $\sigma_{W(\element[][]{H}\alpha)}$ \\\relax
  [JD$-$2\,400\,000 d] & [m~s$^{-1}$] & [m~s$^{-1}$] & [\AA] & [\AA] & [\AA] & [\AA]\\
  \hline
55256.86690 & 2645 & 335 & -6.02 & 0.21 & 9.80 & 0.07\\
55346.70410 & 1593 & 302 & -9.84 & 0.13 & 9.14 & 0.05\\
55347.63521 & 1859 & 315 & -10.44 & 0.17 & 8.54 & 0.06\\
55348.73788 & 2041 & 260 & -13.17 & 0.19 & 9.24 & 0.02\\
55349.74833 & 2223 & 400 & -11.42 & 0.20 & 8.45 & 0.08\\
55351.69950 & 2331 & 112 & -12.80 & 0.24 & 8.45 & 0.03\\
55355.49994 & 2108 & 192 & -8.83 & 0.12 & 8.37 & 0.08\\
55355.68805 & 2300 & 195 & -8.96 & 0.13 & 8.73 & 0.10\\
55355.85300 & 1784 & 784 & -9.85 & 0.20 & 8.84 & 0.11\\
55356.78750 & 1529 & 303 & -8.24 & 0.20 & 9.35 & 0.07\\
55358.63746 & 3925 & 494 & -9.17 & 0.15 & 9.20 & 0.03\\
55358.71724 & 2364 & 408 & -9.15 & 0.14 & 9.21 & 0.06\\
55360.72711 & 2664 & 358 & -10.35 & 0.13 & 8.86 & 0.13\\
55362.48448 & 2432 & 215 & -9.55 & 0.15 & 7.78 & 0.04\\
55362.57222 & 1732 & 205 & -9.67 & 0.11 & 7.97 & 0.07\\
55396.51686 & 3415 & 245 & -10.48 & 0.20 & 7.63 & 0.06\\
55396.57497 & 2641 & 306 & -10.73 & 0.08 & 7.90 & 0.03\\
55396.61357 & 2388 & 123 & -10.91 & 0.17 & 8.37 & 0.02\\
55396.66325 & 3566 & 214 & -10.90 & 0.16 & 8.69 & 0.05\\
55396.69535 & 2672 & 382 & -11.18 & 0.08 & 8.75 & 0.03\\
55397.52834 & 1992 & 339 & -10.01 & 0.11 & 8.27 & 0.05\\
55397.59554 & 2288 & 247 & -9.99 & 0.13 & 8.11 & 0.11\\
55397.64898 & 2739 & 381 & -10.02 & 0.30 & 8.05 & 0.03\\
55397.68781 & 2866 & 241 & -10.27 & 0.15 & 8.06 & 0.09\\
55398.51102 & 1932 & 519 & -11.25 & 0.29 & 8.45 & 0.09\\
55400.51882 & 2692 & 303 & -10.66 & 0.14 & 10.03 & 0.06\\
55400.57368 & 2159 & 173 & -10.18 & 0.22 & 9.65 & 0.06\\
55400.61784 & 2891 & 239 & -9.70 & 0.14 & 9.64 & 0.10\\
55400.67505 & 2803 & 321 & -9.52 & 0.20 & 9.41 & 0.07\\
55401.48617 & 3341 & 376 & -10.65 & 0.17 & 9.20 & 0.03\\
55401.63979 & 3461 & 215 & -10.01 & 0.17 & 9.23 & 0.02\\
55401.69736 & 1035 & 546 & -9.56 & 0.21 & 8.93 & 0.07\\
55402.56980 & 2032 & 374 & -9.41 & 0.00 & 8.63 & 0.06\\
55402.62734 & 3849 & 347 & -9.35 & 0.24 & 8.36 & 0.11\\
55402.69271 & 1769 & 433 & -8.77 & 0.18 & 8.09 & 0.06\\
55403.56511 & 2802 & 378 & -8.49 & 0.19 & 8.81 & 0.04\\
55403.59827 & 2209 & 193 & -8.08 & 0.23 & 8.48 & 0.13\\
55403.63910 & 2810 & 471 & -8.05 & 0.19 & 8.60 & 0.11\\
55403.67692 & 1898 & 311 & -7.99 & 0.23 & 8.59 & 0.11\\
55404.50509 & 3058 & 250 & -8.80 & 0.18 & 9.55 & 0.05\\
55404.67289 & 2932 & 247 & -9.56 & 0.18 & 9.61 & 0.06\\
55405.63121 & 2431 & 197 & -12.65 & 0.16 & 9.74 & 0.07\\
55405.68095 & 2686 & 313 & -12.62 & 0.19 & 9.55 & 0.04\\
55407.58835 & 3106 & 371 & -10.67 & 0.16 & 8.52 & 0.02\\

  \hline
  \end{tabular}
\end{table*}
\end{document}